\documentclass{aa}
\usepackage[usenames,dvipsnames]{xcolor}
\usepackage[T1]{fontenc}
\usepackage{natbib}
\usepackage{csquotes}
\usepackage{xspace}
\usepackage{amssymb}
\usepackage{amsmath}
\usepackage{graphicx}
\usepackage{hyperref}
\usepackage{siunitx}
\usepackage{subcaption}
\usepackage{aas_macros}

\DeclareSIUnit\erg{erg}
\DeclareSIUnit\Myr{Myr}
\DeclareSIUnit\AU{AU}

\def\chemcomp{\texttt{chemcomp}\xspace}

\def\erg{\hbox{erg}}

\def\AU{\hbox{AU}}

\begin{document}

\title{
\chemcomp: Modeling the chemical composition of planets formed in protoplanetary disks
}
\authorrunning{Schneider \& Bitsch}
\author{Aaron David~Schneider$^{1,2}$ \& Bertram~Bitsch$^{3,4}$}
\institute{
  (1) Centre for ExoLife Sciences, Niels Bohr Institute, Øster Voldgade 5, 1350 Copenhagen, Denmark\\
  (2) Instituut voor Sterrenkunde, KU Leuven, Celestijnenlaan 200D, B-3001 Leuven, Belgium \\
  (3) Max-Planck-Institut f\"ur Astronomie, K\"onigstuhl 17, 69117 Heidelberg, Germany\\  
  (4) Department of Physics, University College Cork, Cork, Ireland
} \date{\today}
\offprints{A. Schneider,\\ \email{aarondavid.schneider@kuleuven.be}}
\abstract{
Future observations of exoplanets will hopefully reveal detailed constraints on planetary compositions. Recently, we have developed and introduced \chemcomp \citep{Schneider2021I}, which simulates the formation of planets in viscously evolving protoplanetary disks by the accretion of pebbles and gas. The chemical composition of planetary building blocks (pebbles and gas) is traced by including a physical approach of the evaporation and condensation of volatiles at evaporation lines. We have now open-sourced the \chemcomp code to enable comparisons between planet formation models and observational constraints by the community. The code can be found at \url{https://github.com/AaronDavidSchneider/chemcomp}, is easy to use (using configuration files) and comes with a detailed documentation and examples.
  }

\maketitle
Described in detail in \citet{Schneider2021I}, \chemcomp is capable of modeling the formation of planets in the core-accretion paradigm by the accretion of pebbles and gas, while tracing the chemical composition of the individual building blocks, including the gas and pebble composition. To this date, several works have used the original version of \chemcomp, including  \citet{Bitsch+2021,Schneider2021II,BitschSchneider2022,BitschMah2023a,MahBitsch2023,MahBitsch2023CO,SavvidouBitsch2023a}. Furthermore, several add-ons have been implemented to account for additional physics \citep{HuehnBitsch2023chemcomp, Chatziastros2023, DantiBitsch2023chemcomp}. 

In its core, \chemcomp solves the viscous disk equation and dust transport equation for a set of molecular species in a 1D grid-discretized protoplanetary disk using an adapted version of the donor-cell advection-diffusion solver from the unpublished \texttt{disklab} \citep{disklab} code. Evaporation and condensation of molecular species at icelines is taken into account by source and sink terms in the above mentioned equations to account for the phasetransition that a molecular species undergoes, when crossing its iceline. A planet can be placed at any time into the disk, which will then accrete first pebbles and later gas taking into account migration of type I and II and the formation of a gap in the disk.

To expedite scientific progress, we have provided the original code and documentation from the work of \citet{Schneider2021I} for free use. The code can be accessed at \url{https://github.com/AaronDavidSchneider/chemcomp}. The plugins developed by \citet{HuehnBitsch2023chemcomp, Chatziastros2023, DantiBitsch2023chemcomp} are not part of this version of the code, but will be made available upon reasonable request.

\begin{acknowledgements}
A.D.S and B.B. \ thank the European Research Council (ERC Starting Grant 757448-PAMDORA) for their financial support. A.D.S. acknowledges funding from the European Union H2020-MSCA-ITN-2019 under Grant no. 860470(CHAMELEON) and from the Novo Nordisk Foundation Interdisciplinary Synergy Program grant no. NNF19OC0057374. We thank Cornelis Dullemond for discussions about pebble condensation and for providing the numerical solver for the advection-diffusion equations.
\end{acknowledgements}

\begingroup
\bibliographystyle{aa}
\bibliography{ms}
\endgroup

\end{document}